\documentclass[epj]{svjour}

\usepackage{graphicx}
\usepackage{fancyhdr}
\def\makeheadbox{{
 }}
\def\mytitle{My title} 
\def\myauthors{My name}  
\def\mytype{My type of session}
\def\mysession{My session}

\def\mytitle{Study of DiMuon Rare Beauty Decays with ATLAS and CMS} 
\def\myauthors{A. Policicchio and G. Crosetti}    
\def\mytype{Contributed Talk}    
\def\mysession{Flavor Physics}

\begin{document}
\title{Study of DiMuon Rare Beauty Decays with ATLAS and CMS}
\subtitle{}
\author{A. Policicchio\inst{}
\thanks{\emph{Email:} antonio.policicchio@cern.ch}%
 \and
 G. Crosetti\inst{}
}                     
%
%
\institute{Universita' della Calabria and INFN Cosenza \\ \\ representing the ATLAS and CMS collaborations}
%
\date{}
\abstract{ The LHC experiments will perform sensitive tests of physics
beyond the Standard Model (BSM). The investigation of decays of beauty
hadrons represents an alternative approach in addition to direct BSM
searches. The ATLAS and CMS efforts concentrate on those $B$-decays
that can be efficiently selected already at the first and second level
trigger. The most favorable trigger signature will be for $B$-hadron
decays with muons in the final state. Using this trigger, ATLAS and
CMS will be able to accommodate unprecedentedly high statistics in the
rare decay sector. These are purely dimuon decays, and families of
semimuonic exclusive channels. Already with data corresponding to an
integrated luminosity of \ensuremath{1~fb^{-1}}, the sensitivity in the
dimuon channels will be comparable to present measurements (world
average).  The strategy is to carry on the dimuon channel program up
to nominal LHC luminosity. In particular the \ensuremath{B_s \to
\mu\mu} signal with \ensuremath{\sim}5 sigma significance can be
measured combining low luminosity \ensuremath{10^{33}cm^{-2} s^{-1}}
samples with those of one year of LHC operation at a luminosity of
\ensuremath{10^{34}cm^{-2} s^{-1} }.
} 
\maketitle

\section{Introduction}
\label{intro}
Rare leptonic and semileptonic $B$-decays, produced by FCNC
transitions, are forbidden at the tree level in the Standard Model
(SM). These decays occur at the lowest order only through one-loop
``penguin'' and ``box'' diagrams. The branching ratios of these decays
are very small: from $4 \times 10^{-5}$ for the rare radiative decay
$B^0_d \to K^* \gamma$ to $10^{-15}$ for the rare Cabibbo suppressed
leptonic decay $B^0_d \to e^+ e^-$.

The careful investigation of rare $B$-decays is mandatory for testing
ground of the Standard Model and offers a complementary strategy in the
search of new physics. The probing of loop-induced couplings provide a
means of testing the detailed structure of the SM at the level of
radiative corrections. In particular, FCNC involving $b \to s,d$
transitions and $B \to ll$ decays, provide an excellent probe of new
indirect effects by yielding informations on the masses and coupling
of the virtual particles running in the loops. In SUSY models, the
branching fraction for $B^0_{s(d)} \to \mu^+ \mu^-$ has a strong
dependence on $tan\beta$. A precise measurement of such decays will
allow to constrain the supersymmetric extensions of the SM.

To date the decay modes $B^0_{s(d)} \to \mu^+ \mu^-$ have not yet been
observed. The current best upper limits on the branching ratio come
from the D0 \cite{Dzero} and CDF \cite{Cdf} collaborations and are
$9.3 \times 10^{-8}$ and $5.8 \times 10^{-8}$ respectively at
$95\%CL$. The searches for rare $B$ decays at the $B$-factories CLEO,
Belle and BaBar have no sensitivity to $B_s$ decays.

In the last years the $B$-factories BaBar and Belle presented the
first results for $B \to (K^*,K) l^+ l^-$ branching ratios and
forward-backward asymmetry ($A_{FB}$) in these rare semileptonic
decays \cite{babar},\cite{belle1},\cite{belle2} but those are still
affected by large statistical errors.

In this report we pay attention to the (semi)leptonic decays with
$\mu^+ \mu^-$ pairs in final states where ATLAS and CMS can give a
significant contribution. We discuss the simulation results and the
perspectives of measurements. The SM branching ratios of the decays
studied can be found in Table \ref{tab:decays}.

\begin{table}
  \caption{\it Standard Model branching ratios for rare $B$-decays
    into a $\mu^+ \mu^-$ final state.}
  \label{tab:decays}
  \begin{center}
    \begin{tabular}{lll}
      \hline\noalign{\smallskip}
      Decay channel & Br. ratio & Ref. \\
      \noalign{\smallskip}\hline\noalign{\smallskip}
      $B^0_s \to \mu^+ \mu^-$ & $3.4 \times 10^{-9}$ & \cite{Buras} \\ 
      $B^+ \to K^+ \mu^+ \mu^-$ & $3.5 \times 10^{-7}$ & \cite{Ali1} \\
      $B^+ \to K^{*+} \mu^+ \mu^-$ & $\sim 10^{-6}$ & \cite{Ali1} \\
      $\Lambda _b \to \Lambda \mu^+ \mu^-$ & $2.0 \times 10^{-6}$ & \cite{chen},\cite{aliev} \\ 
      $B^0_d \to K^{0*} \mu^+ \mu^-$ & $1.3 \times 10^{-6}$ & \cite{melikhov1},\cite{melikhov2} \\ 
      $B^0_s \to \phi ~ \mu^+ \mu^-$ & $\sim 10^{-6}$ & \cite{melikhov1},\cite{melikhov2} \\ 
      \noalign{\smallskip}\hline
    \end{tabular}
  \end{center}
\end{table}

\section{Theoretical description}
\label{sec:1}
From the theoretical point of view, the $b \to q$ ($q = s,d$)
transitions are described using the effective Hamiltonian

\begin{equation}
  {\cal{H}}_{eff}=-4 \frac{G_{F}}{\sqrt{2}} V_{tb}V^*_{tq}
  \sum_{i=1}^{10} C_{i}(\mu) O_{i}(\mu)
  \label{effham}
\end{equation}

in the form of Wilson expansion (see e.g. \cite{Buras96}). The set of
Wilson coefficients $C_{i}(\mu)$ depends on the current model and
contains the lowest order model contributions and perturbative QCD
corrections. The scale parameter $\mu$ is approximately equal to the
mass of the $b$-quark ($\sim 5$~GeV). This parameter separates the
perturbative and nonperturbative contributions of the strong
interactions. The nonperturbative contribution is contained in the
matrix elements of basis operators $O_{i}(\mu)$ between the initial
and final hadronic states. For the calculation of these matrix
elements it is necessary to use different decay-specific
nonperturbative methods: QCD Sum Rules, Heavy Quarks Effective Theory,
Quark Models and Lattice calculations.  The accuracy in
nonperturbative calculations depends on the method, but it is not less
than 15$\%$. The accuracy of the Wilson coefficient with NLO and NNLO
QCD corrections \cite{bobeth} is not greater than 15$\%$ if the $\mu$
parameter ranges in $[m_b/2,2m_b]$.

In the SM the decay width of the rare muonic decays is:

\begin{eqnarray}
  \Gamma(B^0_q \to \mu^+ \mu^-)=\frac{G_{F}^{2}\alpha_{em}^{2}}{16\pi^3} |V_{tq}^{*}V_{tb}|^2 \cdot
  \\\Big(f_{B_q} m_{\mu} C_{10A}\Big)^2\sqrt{M^{2}_{B_q} - 4m^2_{\mu}}.\nonumber
  \label{decwidth}
\end{eqnarray}

This expression contains only one nonperturbative constant
$f_{B_q}$. The value of this constant is known with an accuracy of the
order of about 5-10$\%$. Furthermore the Wilson Coefficient $C_{10A}$
in the NLO approach is not dependent on the scale parameter $\mu$, and
does not add any uncertainty to the theoretical predictions.

\section{Trigger strategies for rare decays}
\label{sec:2}
Details of the ATLAS and CMS experiments can be found in
Refs. \cite{ATLASTRD,CMSTDR}.

\subsection{The ATLAS trigger}
\label{subsec:21}
ATLAS has a three level trigger system \cite{ATLASTRIG} which reduces
the 40~MHz bunch crossing rate to about 100~Hz of events to be
recorded. The first level trigger (LVL1) is hardware-based and makes a
fast decision (in 2.5~$\mu$s) about which events are interesting for
further processing. Coarse granularity informations from calorimeter
and muon spectrometer are used to identify region of interest (RoI) of
the detector which contain interesting signals (high energy electrons,
muons and taus and jets with large transverse or missing energy). The
RoIs are used to guide the later stages of the trigger. After LVL1 the
trigger rate will be reduced to less than 75~kHz.

The high level trigger (HLT) is software-based and is split into two
levels. At the level 2 (LVL2) the full granularity of the detector is
used to confirm the LVL1 decisions and then to combine informations
from different sub-detectors within the LVL1 RoIs. Fast algorithms are
used for the reconstruction at this stage and the rate is reduced to
$\sim$2~kHz with an average time of execution of $\sim$10~ms. At the
level 3, the Event Filter (EF), the whole event is available and
offline-like algorithms are used with better alignment and calibration
informations to form the final decision. The rate is reduced to 100~Hz
with an execution time of $\sim$1~s.

\subsubsection{ATLAS trigger for rare decays}
The $B$-trigger is expected to account for 5-10$\%$ of the total
trigger resources. The core of the $B$-trigger is the LVL1 muon
trigger which is based on the measurement of the muon transverse
momentum ($p_T$). 
The efficiency of the muon trigger is expected at about 85$\%$ at the
plateau. The dimuon LVL1 trigger (two muons with $p_T$ above 6~GeV),
used for rare decay selection, is expected to have a rate of about
500~Hz.

The LVL1 dimuons will be confirmed at the LVL2 firstly in the muon
system by means of the precision tracking chamber and then by
combining muon and inner detector tracks. Finally the two muons can be
combined and mass cuts are applied. At the EF the tracks are refitted
in the RoIs and vertex reconstruction is performed. Cuts are applied
on decay length and invariant $B$-hadron mass.  For $B^0_s \to \mu^+
\mu^-$ events containing two muons with $p_T >$6~GeV, efficiencies of
60-70$\%$ are expected.

\subsection{The CMS trigger}
\label{subsec:22}
CMS has a two level trigger \cite{CMSTDR} which reduces the bunch
crossing rate down to about 150Hz for recording. The Level-1 (L1)
trigger uses muon detector and calorimeter informations and is
hardware-based with an output rate of about 100~kHz and a latency of
3.5~$\mu$s. The HLT is software-based with the required 150~Hz output
rate. It uses reconstruction algorithms similar to the offline with a
mean execution time per event of about 40~ms.  To speed up
reconstruction in the HLT, a partial track reconstruction is
performed: the track resolution becomes asymptotic after 5-6 hits are
used in the track fit.

\subsubsection{CMS trigger for rare decays}
As in ATLAS, the CMS trigger for $B$ events uses single and dimuon
triggers. The L1 dimuon trigger has a low $p_T$ threshold of 3~GeV
which ensures a high efficiency for events with two muons in the final
state with a rate of 0.9~kHz at $2 \times 10^{33}$~cm$^{-2}$s$^{-1}$.
At the HLT the L1 decision will be confirmed using the full muon
system and an improved momentum measurement with the tracker. Primary
vertex (PV) reconstruction is available from the pixel detector and
also track reconstruction is performed in cones around the L1 muons
using the partial track reconstruction. The exclusive rare decay is
then reconstructed and cuts on the invariant mass, vertex fit quality
and decay length are applied.

\section{Muonic decays in ATLAS and CMS}
\label{sec:3}
As purely leptonic $B$-decays are theoretically very clean, they
provide an ideal channel for seeking indirect hints of new physics
effects. However, they are very difficult to observe because of their
small branching ratio (see Table \ref{tab:decays}). Most probably they
will not be observed by other experiments before the LHC data
taking. ATLAS and CMS will start sensitive measurements at
$10^{33}$~cm$^{-2}$s$^{-1}$. Even at high design luminosity
($10^{34}$~cm$^{-2}$s$^{-1}$) the trigger for $B \to \mu^+ \mu^-$ decay
is not problematic.

A good background rejection is necessary for the signal selection. The
main contributions to the background \cite{ATLASBG} come from the
processes $b \bar b (b \bar b b \bar b, b \bar b c \bar c) \to X \mu^+
\mu^-$ with the muons originating mainly from semi- \\ leptonic $b$
and $c$ quark decays. This background can be estimated by
extrapolating the Tevatron data on heavy quark production to the LHC
energies. It would be remarked that both ATLAS and CMS analysis are
limited by the statistics of the MonteCarlo background sample.

The event selection relies on topological variables related to the PV,
the muon candidates and the $B_s$ secondary vertex and is very similar
in both experiments \cite{ATLASBMUMU,CMSBMUMU}.  Simple cuts can
be applied to distinguish the combinatorial background from the signal:

\begin{itemize}
\item $B$-hadron invariant mass;
\item secondary vertex length and quality;
\item pointing of $B$-hadron momentum to PV;
\item track isolation.
\end{itemize}


Table \ref{tab:masses} summarizes the mass resolution and the proper
time resolution obtained on the signal MonteCarlo event sample for CMS
and ATLAS.

\begin{table}
  \caption{\it Mass and proper time resolution obtained on the $B^0_s
\to \mu^+ \mu^-$ signal MonteCarlo event sample for CMS and ATLAS.}
  \label{tab:masses}
  \begin{center}
    \begin{tabular}{lll}
      \hline\noalign{\smallskip}
      Experiment & Mass res.(GeV) & Proper time res.(fs)\\
      \noalign{\smallskip}\hline\noalign{\smallskip}
      ATLAS & 0.084 & 91\\ 
      CMS & 0.036 & 95\\
      \noalign{\smallskip}\hline
    \end{tabular}
  \end{center}
\end{table}

The expected number of signal and background events after 30~fb$^{-1}$
are summarized in Table \ref{tab:upperl}. This will allow to set a
stringent constraint on new physics models.

\begin{table}
  \caption{\it Expected number of $B^0_s \to \mu^+ \mu^-$ signal and
  background events after 30~fb$^{-1}$.}
  \label{tab:upperl}
  \begin{center}
    \begin{tabular}{lll}
      \hline\noalign{\smallskip}
      Experiment & signal events & BG events \\
      \noalign{\smallskip}\hline\noalign{\smallskip}
      ATLAS & 21.0 & 60$\pm$36 \\ 
      CMS & 18.3 & $42^{+66}_{-42}$ \\
      \noalign{\smallskip}\hline
    \end{tabular}
  \end{center}
\end{table}

\begin{figure}
\includegraphics[width=0.45\textwidth,height=0.45\textwidth,angle=0]{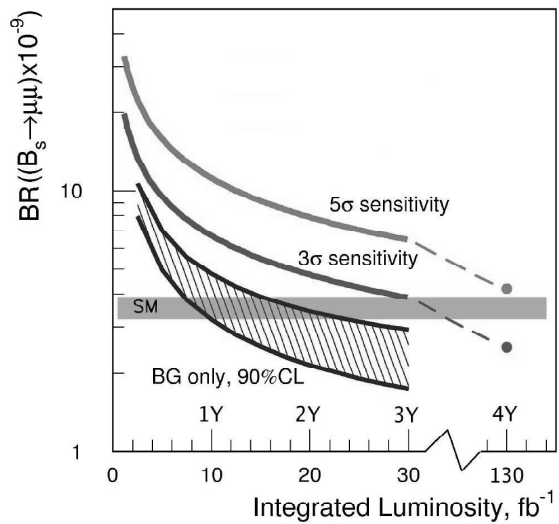}
\caption{ATLAS perspective for the measurement of the $B^0_s \to \mu^+
\mu^-$ branching ratio as a function of the integrated luminosity (or
equivalently as a function of the time). The shaded band shows the
uncertainty in the BG level estimation.}
\label{fig:limits}      
\end{figure}

Figure \ref{fig:limits} shows the ATLAS expectation for the measurement
of the $B^0_s \to \mu^+ \mu^-$ branching ratio as a function of the
integrated luminosity (or equivalently as a function of time). The
SM expectation can be reached with a $\sim$5 sigma significance
combining low luminosity $10^{33}$~cm$^{-2}$s$^{-1}$ samples with those
of one year of LHC operation at the nominal luminosity of
$10^{34}$~cm$^{-2}$s$^{-1}$.

\section{SemiMuonic decays in ATLAS}
\label{sec:4}
Thanks to the dimuon final state (see Table \ref{tab:decays}), the
semimuonic decays, as the purely dimuonic ones, are easy to select at
the trigger level. The observation of semileptonic decays give access
to a number of observables. The precise measurements of such
observables could give very interesting informations for new physics
reach. $A_{FB}$ is one of the most promising parameters.

The small branching ratios of semimuonic decays require a powerful
background rejection.  Semileptonic decays with $c \bar c$ resonances
decaying into two muons represent an irreducible background source. A
cut on the dimuon invariant mass around the nominal values for
resonances removes this background. Combinatorial background arises
from muons originating mainly from semi- leptonic decays of $b$ and
$c$ quarks. Specific decay channels can represent background sources
due mainly to hadron misidentification as muons, but their
contribution is expected to be less important than the previous two
sources.

The event selection \cite{POLIX1,POLIX2} is related to topological
variables:

\begin{itemize}
\item vertex quality and invariant mass of the dimuon system;
\item displacement and quality of vertices and mass of the secondary
hadrons;
\item pointing of $B$ hadron momentum to PV.
\end{itemize}

The number of events expected after three years of data taking at low
luminosity (30~fb$^{-1}$) are summarized in Table
\ref{tab:expected}. The background level estimation is only limited by
the low Monte Carlo statistics available at the moment. It should be
pointed out that, thanks to the muon pair in the final state,
semimuonic decays will be also studied at high luminosity, so that a
larger statistics than those of Table \ref{tab:expected} will be
collected.

\begin{table}
  \caption{\it Expected number of events from semimuonic decays events
  and expected background events in ATLAS after 30~fb$^{-1}$.}
  \label{tab:expected}
  \begin{center}
    \begin{tabular}{lll}
      \hline\noalign{\smallskip}
      Decay & Signal events & Background events\\
      \noalign{\smallskip}\hline\noalign{\smallskip}
      $B^+ \to K^+ \mu^+ \mu^-$ & 4000 & $<$10000 \\
      $B^+ \to K^{*+} \mu^+ \mu^-$ & 2300 & $<$10000 \\
      $\Lambda_b \to \Lambda \mu^+ \mu^-$ & 800 & $<$4000 \\ 
      $B^0_d \to K^{0*} \mu^+ \mu^-$ & 2500 & $<$10000 \\ 
      $B^0_s \to \phi ~ \mu^+ \mu^-$ & 900 & $<$10000 \\     
      \noalign{\smallskip}\hline
    \end{tabular}
  \end{center}
\end{table}

\begin{figure}
\includegraphics[width=0.45\textwidth,height=0.45\textwidth,angle=0]{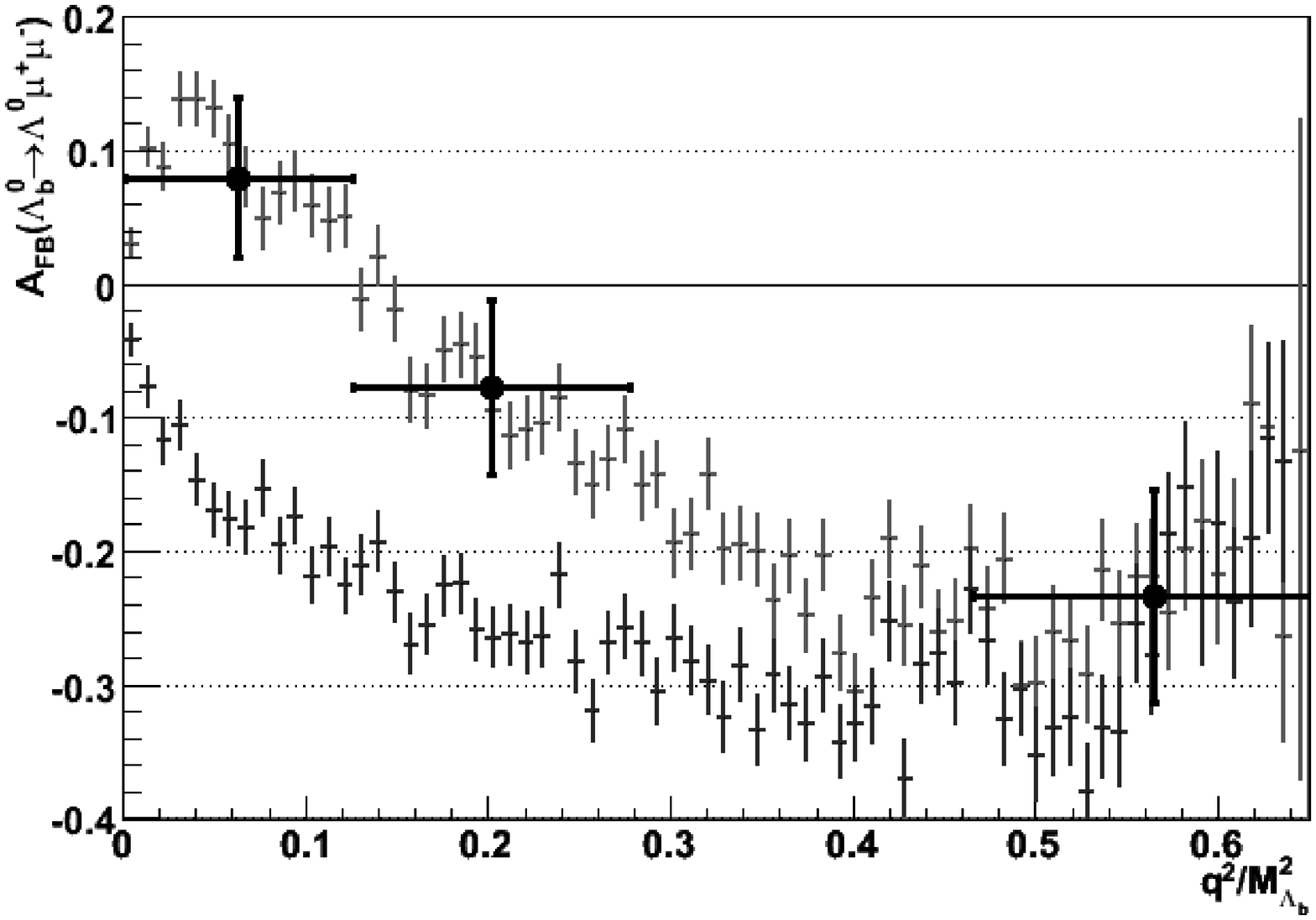}
\caption{Forward-backward asymmetry for $\Lambda_b \to \Lambda \mu^+
\mu^-$ as a function of the dimuon invariant mass after 30f~b$^{-1}$
(see Section \ref{sec:4}).}
\label{fig:asymm}      
\end{figure}

The expected precision on $A_{FB}$ after 30~fb$^{-1}$ is presented on
Figure \ref{fig:asymm} for $\Lambda_b \to \Lambda \mu^+ \mu^-$
decay. The three dots with error bars correspond to simulated data
after offline analysis. The upper point set corresponds to the
theoretical SM prediction, and the lower set corresponds to a
prediction within the MSSM \cite{chen}. The statistical error in the low dimuon
invariant mass region is at level of 6$\%$. The statistical errors
expected on branching ratio measurements are at the level of 3.5$\%$
and 6.5$\%$ respectively for $B \to K \mu^+ \mu^-$ and $B \to K^{*}
\mu^+ \mu^-$ decays. These errors on the branching ratio measurements
are much smaller than the current experimental and theoretical ones.

\section{Conclusions}
\label{sec:5}
The results obtained for $B^0_s \to \mu^+ \mu^-$ by ATLAS and CMS are
comparable and promise an interesting startup analysis with the
possibility of setting tight constraints on new physics models beyond
the SM. The simulation studies show that the ATLAS detector will be
capable to extract signals of semimuonic $B$-decays and reach a good
sensitivity to new physics beyond the SM.

\section{Acknowledgments}
\label{sec6}
We would thank ATLAS and CMS Collaborations, and especially Maria
Smizanska, Sergey Sivoklokov, Pavel Reznicek from ATLAS and Christina
Eggel and Urs Langenegger from CMS for their useful suggestions in the
preparation of this work.


\end{document}